\documentclass[12pt]{article}
\newcommand{\g}{\gamma}

\newcommand{\grad}{{\rm grad}}

\newcommand{\e}{\epsilon}

\newcommand{\ar}{\longrightarrow}

\newcommand{\la}{\lambda}
\renewcommand{\a}{\alpha}
\begin{document}
\title{Genetic simulation of quantum dynamics by the principle of quantum state selection}
\author{Yu.I.Ozhigov \thanks{The work is supported by Fond of NIX Computer Company grant \# F793/8-05, INTAS grant 04-77-7289 and Russian Foundation for Basic Research grant 06-01-00494-a. } \\[7mm]
Moscow State University,\\
chair of quantum informatics, MSU, VMK, Vorobievi Gori, \\GSP-1,119992, Moscow, Russia 
} 
\maketitle
 PACs: 07.05.Tp, 03.67.Lx


\begin{abstract}
The simple genetic algorithm is proposed for the simulation of quantum many body dynamics. It uses the selection of entangled quantum states and has the inbuilt absolute decoherence that comes from the limitation of classical memory. It utilizes the "pre-quantum field" in the form of interacting between the different "quantum worlds". It is shown how this selection model can be applied to the problem of molecular association in chemical reactions.  
\end{abstract}

\section{Introduction and background}

Algorithmic approach to quantum theory was proposed by the author in previous works (see \cite{3}, \cite{4}). It is based on our firm conviction that the effective classical algorithms with expert estimations of a user represent the sufficient tool for the complete description of the Nature on any levels including quantum theory. The key requirement in this approach is the simplicity and the transparency of algorithms which are designed for the simulation of main processes in the micro-world. The expected advantage of algorithmic approach comes from its possibility to give the effective algorithms in cases when the standard quantum theory gives no algorithms for many of such processes, and just here we can expect the advantage of algorithmic approach over the standard Copenhagen quantum theory. 

The first example can come from chemistry, where the application of quantum methods has the long history. Nevertheless, quantum chemistry takes up stationary electronic configurations, conformations of molecules and bound energies only, and there is no robust chemical simulator. This situation is not random, because there is no full quantum description of real dynamical processes. Shroedinger equation is not applicable even to the simplest chemical reaction like the capture of a free electron by the Coulomb field of a proton. Here the probability to obtain the electron in state 1s does not change in time that makes the capture impossible. In the reality the emission of free photons always plays the key role in the reactions of association of molecules. Just the emitted photons take off the energy for the moving atoms that makes possible their joint in the molecule. Unfortunately, quantum electrodynamics (QED) gives us no robust algorithms for such processes as well. QED leads to the divergence of sums for amplitudes and there is no completely satisfactory method to avoid it. Even if such a method is found QED is not applicable to chemistry due to the more fundamental reason than the divergence of sums. QED is the part of quantum theory and hence it inherits its basic drawback: the principal absence of the integral description of quantum evolution. Unitary dynamics in quantum theory is strictly separated from the measurements, and quantum theory factually is the theory of unitary dynamics whereas for the full description of chemistry we manage with the both these types of dynamics. The account of photons in chemical reactions has no sense without some certain supposition about when the collapse of quantum state vector happens. Just this supposition lies beyond the framework of quantum theory, and this deprives the models of chemical reactions the status of exact theory. 

The wish to describe chemistry leads us to the necessity to extend the methods of quantum theory to the dynamics of many particle processes. My opinion is that such an extension means the modification of the mathematical apparatus of quantum theory, namely we must use algorithmic approach instead of analytic and algebraic methods (see the previous discussion in \cite{7}). Only this radical step makes possible to build the consistent description of chemical reactions.

The reason that the robust description of chemical reactions must unavoidably have quantum character lies in the fact that the essence of these reactions has quantum nature, namely, it is based on the fundamental notion of entangled states. In this paper I represent the model of chemistry where the basic element is quantum entanglement between the particles participating in the reaction. In general sense, this gives the new argument for algorithmic approach. In practical sense – for those, who are looking for the robust algorithms – this gives the good starting point for elaboration of effective algorithms for the simulating of chemical reactions. 

\section{Method of collective behavior}

The method of collective behavior represents the good alternative for the algebraic description of quantum one particle evolution. The discussion about the previous versions of this approach (Bohm method) can be found in \cite{7}, cite{11}. The conventional matrix algebra leads to the tremendous non effective usage of computational resources because of the following reason. When calculating the resulting matrix of quantum unitary evolution for time segment consisting of two parts: from $t_0$ till $t_1$ and from $t_1$ till $t_2$, we have to multiply these matrices. The multiplication factually means that we trace on all possible paths including those where no particle exists in order only to convince that there is no particle there! Factually, the bulk of quantum interference has the destructive character that is the reason of the non effective usage of classical computational time. 

Instead of matrix algebra, we represent one particle not by its wave function, but as the ensemble 

\begin{equation}
S=\{ s_1,s_2,\ldots,s_m\}
\end{equation}
of classical particles which are treated as the samples of the real quantum particle. We call this ensemble the swarm. The main property of a swarm is 
that the density of samples determined as 
$$
\rho(\bar r)=\lim\limits_{dx\ar\infty}\frac{N_{\bar r,\ dx}}{(dx)^3},
$$
where $ N_{\bar r,\ dx}$ is the total number of the samples in the cube with the side $dx$ which center is $\bar r$. That is Born rule for quantum probability is guarantied for the natural treatment of the probability, which does not use the observation. This natural treatment is based on the classical urn scheme for the choice of sample that determines the position of the real quantum particle. We then exclude the notion of observation from the kit of formal tools, and treat observations as the special kind of quantum evolution. Fortunately, it is possible to reduce quantum dynamics to the simple interactions between samples in the swarm; this reduction factually includes decoherence and can be treated as the universal mean for the integral description of quantum dynamics without the division to unitary dynamics and measurements.

The correspondence between swarm representation of the quantum particle and its standard representation through the wave function is given by the following formulas for the density of samples and total impulse of the swarm in the small cube around the point $r$: $\bar p(r)$ and the relative phase of wave function $\phi$ ($\Phi=\sqrt{\rho }e^{i\phi}$, so that the phase in the point $r_0$ is zero), where $\bar v=\bar p/\rho$ is the mean velocity of samples in this cube:
\begin{equation}
\begin{array}{ll}
&|\Psi(r)|=\sqrt{\rho(r)};\\
&\phi(r)=\int\limits_{\g :\ r_0\ar r}k(dx)^2\bar v\cdot d\g,\\
&\bar v = a(dx)^{-2}\grad\ \phi(r),
\end{array}
\label{connection}
\end{equation}
where $\cdot$ denotes scalar product of vectors.
These formulas determine the passage from swarm representation of quantum dynamics to the wave function and vice versa provided the grain of spatial resolution $dx$ is fixed. The quantum behavior of real particle can be easily approximated with any desired accuracy by the swarm with simple rules of behavior for its members. This rule is: the exchange of impulses between close samples. It is proved in \cite{4} that in this case the simple mechanism of impulse exchange ensures the approximate coincidence of these two descriptions of quantum dynamics, within $dx^3$ in the determining of the wave function.

The rule of impulse exchange is not rule of classical physics though it conserves the total impulse of the swarm. The interaction with the external field is classical for all the samples. We do not take up the question about what does transmit the impulse from one sample to the other and vice versa. It can be shown that this way gives us the approximation of Shroedinger equation if we fix the grain $dx$ of spatial resolution; the intensity of impulse exchange must have the order $(dx)^{-3}$. This peculiarity distinguishes the method of collective behavior from the Bohm mechanics (see cite{11}). The last gives the model for all $dx$ at a time, but has no so transparent rule for microscopic behavior of samples. Just the simple rule for microscopic behavior is needed for the extension of the method of collective behavior to the many particle case. Let us now take up this extension with the method of collective behavior with impulse exchange. 

Let we are given a set of $n$ quantum particles that we enumerate by integers: $1,2,\ldots,n$. We assume that the main act of evolution is the reaction of scattering when these particles fly to each other simultaneously and can associate in some stable complex objects called molecules. Factually, the more general picture of scattering takes place: initial particles can consist of some more elementary particles, and in the reaction these more elementary particles can regroup and form the products of the reactions, which consist of the same elementary particles as the initial objects, but in the other configurations. 

The simple example of such a reaction is the scattering of a proton on an atom of hydrogen. Here the moving proton (proton number one) flies to the staying hydrogen atom which in turn consists of an electron and proton number two. The possible products are: a) isolated proton number one and hydrogen atom (no has happened), b) isolated proton number two and hydrogen atom formed by the electron and the proton number one (the flying proton tears the electron from the staying atom – this reaction is called the recharge), c) forming of the molecular ion of hydrogen (two protons glued by the electron), and d) separate protons and electrons. The cases b) and c) represent the main interest here because we then have the recombination of constituents (b) or association of the new molecule (c). 

We assume that the right description of the elementary reactions with $n$ particles is sufficient to build the actual model of the processes of any degree of complexity including the description of the simple forms of living entities, like viruses and bacteria. Namely, the easy generalization of abstract methods for scattering will give us the picture of the behavior of very complex objects. 

The main requirement to these types of models is that the required time and memory for the simulation must grow not faster than linearly of the total number $n$ of participating particles which is treated as non dividable\footnote{We also assume that the observed chemical dynamics does not depend on the sub nuclear states, e.g. the chemistry is determined by the electrons and nuclei. It is under question for very complex objects but is likely true for not long time frames even for them.} We also assume that our simulation must be based on quantum mechanics, and the single essentially not conventional procedure in the simulation is the simulation of decoherence. 

The easiest algorithmic model of decoherence is called the absolute decoherence model. It claims that decoherence comes as the reduction of quantum state 

\begin{equation}
|\Psi\rangle =\sum\limits_j\la_j|j\rangle
\label{psi}
\end{equation}
in the instant when the memory of the simulating computer cannot include the whole notation of this state. The absolute model can be concretized as follows. We suppose that the amplitudes in \ref{psi} cannot exceed some level $\e>0$, called amplitude grain. If in the unitary evolution some amplitude $\la_j$ becomes less than $\e$, the corresponding summand $\la_j|\j\rangle$ is merely excluded from the state \ref{psi}, with the corresponding renormalization of state. In the work \cite{3} it was shown that this simple rule gives the Born rule for probability to obtain the state $|j\rangle$ as the result of measurement of $\|psi\rangle$ as $p_j=|\la_j|^2$. But this is yet not the final form of simulating algorithm because the rule of small amplitude reduction requires matrix algebra technique and thus cannot serve as the core of simulating algorithms due to the non economical essence of matrix computations. 

To obtain the robust scheme of simulating algorithm we must sequentially use the method of collective behavior, where the algorithmic reduction of quantum state is the inbuilt property. Let us consider the swarm representation of our $n$ particles $1,2,\ldots,n$, where $S_1,S_2,\ldots,S_n$ are the swarms of samples corresponding to their states $|\Psi_1\rangle, |\Psi_2\rangle, \ldots, |\Psi_n\rangle$. If we consider the ensemble consisting of all these samples, it will be the representation of non entangled state of the form $|\Psi_1\rangle\bigotimes |\Psi_2\rangle\bigotimes \ldots\bigotimes |\Psi_n\rangle$. But to represent the entangled state of the form
\begin{equation}
\Phi\rangle = \sum\limits_{j_1,j_2,\ldots,j_n} \la_{j_1,j_2,\ldots,j_n}|j_1,j_2,\ldots,j_n\rangle
\label{many}
\end{equation}
we must introduce the new and crucial element to the method of collective behavior. This is the bonds between the samples of the different swarms. The basic state $j_i$ can be treated as the coordinate of particle $i$ in the corresponding configuration space. The representation of wave function in the form \ref{many} means that there are bonds connecting points $j_1, j_2,\ldots,j_n$ in one cortege. The relative quantity of bonds of this form (their total number divided to the total number of all bonds) is $|\la_{j_1,j_2,\ldots,j_n }|^2$. 

We assume that the bonds connect not spatial points but the samples of real particles. They have the form of corteges
\begin{equation}
\bar s =(s_1,s_2,\ldots,s_n)
\label{cortege}
\end{equation}

where for any $j=1,2,\ldots,n\ s_j\in S_j$. The wave function $|\Phi\rangle$ is then represented by the set $\bar S$ of corteges $\bar s$ so that for each $j=1,2,\ldots,\ s_j\in S_j$ there exist exactly one cortege of the form \ref{cortege}. Each cortege plays the role of the so called world in the many world interpretation of quantum theory. We treat this cortege \ref{cortege} as one probe representation of the $n$ particle system and all interactions goes inside the same cortege whereas the real system state results from the interference of amplitudes corresponding to all cortege which occur in the same spatial cell. We call $\bar S$ the swarm for $n$ particle system.

The density of the swarm $\bar S$ is defined as 
\begin{equation}
\rho_{\bar S}( r_1,r_2,\dots,r_n)=\lim\limits_{dx\ar\infty}\frac{N_{ r_1,r_2,\dots,r_n,\ dx}}{(dx)^{3n}},
\label{density}
\end{equation}
where $ N_{ r_1,r_2,\dots,r_n,\ dx}$ is the total number of cortege which occur in the $3n$ dimensional cube with the side $dx$ and the center $ r_1,r_2,\dots,r_n$.

If the wave function $|\Phi\rangle$ is the tensor product of one particle wave functions:
$$
\Phi\rangle = \bigotimes\limits_{i=1}^{n}|\phi_i\rangle 
$$
The corresponding bonds can be obtained by random choice of samples $s_j\in S_j$ for each $j=1,2,\ldots,n$, forming one cortege $s_1,s_2,\ldots,s_n$.
With this choice of cortege we obtain that the density of swarm satisfy the Born condition which can be written for swarms in the form 
\begin{equation}
\sum\limits_{\bar r\in D}|\langle\bar r|\Phi\rangle |^2 = \frac{N_{\bar r, \bar S}}{N}
\label{born}
\end{equation}
where $D\subset R^{3n}$, $ N_{\bar r, \bar S}$ is the total number of corteges occur in the area $D$. But for the entangled state $|\Phi\rangle$ this choice of corteges for the kit $\bar S$ will not give us the equality \ref{born}. We thus must take \ref{born} as the definition of the choice of corteges in $\bar S$. But to define the swarm we also need the velocities for all samples, namely, we need the generalization of equation \ref{connection} to the case of $n$ real quantum particles.  

Let $\Psi(r_1,r_2,\ldots,r_n)$ be the wave function of $n$ particle system, $\Psi=|\Psi|exp(i\phi(r_1,r_2,\ldots,r_n)$ be its Euler expansion. We denote by $\grad_j\phi(r_1,r_2,\ldots,r_n)$ the gradient of $\Psi$ taken on the coordinates of particle $j$, where $j\in\{ 1,2,\ldots,n\}$ is the fixed integer. The generalization of formulas \ref{connection} to the $n$ particle case has the form
\begin{equation}
\begin{array}{ll}
&|\Psi(\bar r)|=\sqrt{\rho(\bar r)};\\
&\phi(r)=\int\limits_{\bar \g :\ \bar r_0\ar \bar r}k(dx)^2\bar v\cdot d\g,\\
&\bar v = a(dx)^{-2}\bar \grad\ \phi(\bar r),
\end{array}
\label{connection_n}
\end{equation}
where $\bar r$ means $r_1,r_2,\ldots,r_n$, $\bar\grad$ means $\grad_1,\grad_2,\ldots,\grad_n$, and $\bar \g$ is the path in $3n$ dimensional space. The rules \ref{connection_n} is sufficient to determine the swarm given the wave function, if we agree to join the samples into corteges independently of their velocities.
The microscopic mechanism of swarm dynamics takes the following form. Impulse exchange between two corteges of samples: $\bar s=(s_1,s_2,\ldots,s_j,\ldots,s_n)$ and $\bar s’=(s’_1,s’_2,\ldots,s’_j,\ldots,s’_n)$ is impulse exchange between the two samples $s_j$ and $s’_j$ provided $\bar s$ and $\bar s’$ belong to the same spatial cube in the configuration space $R^{3n}$ for $n$ particles. With this definition the reasoning from the paper \cite{3} can be repeated straightforwardly and we obtain that this microscopic mechanism of impulse exchange for $n$ particles ensures the approximation of $n$ particle quantum dynamics within the accuracy of the order $dx^{3n}$ in the determining of wave function. 

The described method of collective behavior gives us the good framework for the economical simulation of quantum evolution. 

\section{Genetic method of entangling} 

The method of collective behavior yet does not give us the algorithm for simulation of quantum system dynamics, because the starting point of this method requires the wave function description. To make the collective behavior method complete we must point how to perform entangling, that is how to choose the initial corteges of samples. This choice must guarantee the best approximation of wave function by corteges \ref{connection_n}. We then can treat one sample as a currier of amplitude grain in sense of \cite{3}, and the swarm dynamics will thus give us the approximation of unitary dynamics and decoherence simultaneously. The task of choosing corteges is thus the core of quantum simulation. 

The experiments in the real simulation show that the task of choice of corteges can hardly be solved by the one step procedure. I suggest the following simple genetic algorithm for finding the corteges, which uses the sequential repetitions of dynamical scenarios when the choice of initial conditions for each repetition will use the result of the previous one. 

We will describe the genetic entangling on the example of scattering of $n$ quantum particles. We start from the non entangled state of them where the particle $j$ state is determined by the wave function $\Psi_j$, or, in swarm representation, by the swarm $S_j$. At the first scenario to determine the initial state of our swarm $\bar S_{ini}$ we choose the corteges $\bar s$ at random. After the fixed small time of evolution $\Delta t$ of the swarm we obtain its final state $\bar S_{fin}$. If we have a huge total number $N$ of samples beforehand, in the swarms $S_j$, we would obtain the good approximation of wave function by $\bar S_{fin}$, where the final quantity of all samples will serve as the decoherence factor. The problem is to use the strictly limited number $N$ of the samples to simulate the real dynamics with the admissible accuracy. Here in case of scattering under admissible accuracy we mean the right separation of the products of reactions: for chemical reactions there is the list of possible products with the corresponding probabilities depending on the initial state of reagents. With this limitation of $N$ we must charge the samples with the two roles: the first is to simulate the unitary dynamics of the wave function, and the second is to simulate the decoherence resulting from the amplitude grain. We note that these two role are not in full agreement with each other. The approximation of the wave function requires the small distance $\|\Psi_{Shoedinger}-\Psi_{swarm}\|$ whereas the decoherence resulting from the amplitude grain nulls all states with the amplitude module less than $\e$ that can give the big discrepancy with the wave function in the unitary evolution especially when the dispersion of amplitude distribution is large. 

We thus have to choose corteges $\bar s$ so that the distribution of samples among them be the most economical for unitary dynamics as well as for decoherence – on the short time segment $\Delta t$. Call the space $R^{6n}$ double configuration space for $n$ particles. The sense of it is that we will consider the pair of states: initial and final. For each cortege $\bar s_{ini}$ in the initial swarm there is one and only one cortege $\bar s_{fin}$ which results from $\bar s_{ini}$ in the swarm evolution. We choose the division of double configuration space for $n$ particles to the cells of the form of cubes, and group the resulting pairs $(\bar s_{ini},\bar s_{fin})$ of corteges into groups ${\cal G}_1, {\cal G}_2,\ldots, {\cal G}_k$ so that each group consists of all pairs which occur in the same spatial cube of the division. 
Let the numbers of elements ${\cal N}_j$ in these groups be ordered directly: ${\cal N}_1\geq {\cal N}_2\geq\ldots\geq {\cal N}_k$. We choose the first $k_1<k$ groups and call the pairs in them right pairs. The other pairs are called wrong. We are now ready to obtain the initial state for the next repetition of scattering. We exclude the wrong pairs from $\bar S$ and reorganize their members to the other corteges by the rules of genetic algorithms. Here the application of various genetic methods like the cross over is appropriate, when we group the former members of wrong corteges likely to the grouping of right corteges. For thus created the second version $\bar S_2$ of initial conditions we launch the repetition again and so on. It results in the sequence of pairs 
\begin{equation}
(\bar S^1_{ini},\bar S^1_{fin}),(\bar S^2_{ini},\bar S^2_{fin}),\ldots
\label{chain}
\end{equation}
where each pair $(\bar S^j_{ini},\bar S^j_{fin})$ represent the digest of the repetition number $j$. The passage from one pair to the next consists of three steps: swarm evolution via impulse exchange, selection and the replication of right pairs. The exchange of impulses between the different corteges plays the role of mutations on the evolutionary programming. If we consider the swarm $\bar S$ as the ‘world’ in the many world interpretation of quantum theory, the impulse exchange between the two corteges means the interaction between the different worlds. The chain \ref{chain} must be abrupt when the number of elements of the selected groups becomes stable. 

We represent the argument for that the method of state selection lies along the core of standard quantum unitary dynamics for many particles. Let us turn to the Feynman path integrals (see \cite{10}) where the wave function $\Psi$ in each time instant $t$ is determined by the following equation:
\begin{equation}
\Psi(t,\bar r)=\int\limits_{R^{3n}}K(t,\bar r, t_1,\bar r_1)\Psi(t_1,\bar r_1),
\end{equation}
where $K$ is the kernel of our system, that can be treated as the amplitude careered by a cortege, if we assume that the samples carrier complex numbers – amplitudes instead of their velocities as in the collective behavior method. Amplitudes $K$ careered by a cortege, thus depend on the initial and the final position $\bar r_1$ and $\bar r$, and must be then closed for the corteges which initial and final spatial positions are closed. Let us estimate the deposit to the probability $|\Psi(t,\bar r)|^2$ of two groups of corteges with $l$ elements each: the first group contains in the same group ${\cal G}_j$, and the corteges from the second group have only the final positions closed but the initial are different, and, for simplicity, randomly chosen. The deposits of these two groups to the probability are approximately
$$
d_1=|\sum\limits_{s=1}^l\a |^2=k^2|\a |^2,\ \ d_2=|\sum\limits_{s=1}^l\a e^{i\phi_s}\approx k|\a |^2
$$
where the phases $\phi_s$ are distributed randomly. The last approximate equality follows from the fact that for the uniform distribution of $\phi_s$ the medium of the distance of this sum from zero must have the order square root of the module of a summand. Hence, the deposit of the first group is prevailing. Returning to the samples in the collective behavior, we note that the complex amplitudes here are substantiated by the velocities of samples, and the deposit of minor groups ${\cal G}_j$ for $j> k_1$ will be much smaller than the deposit of the right samples taken in the same quantity, that legalizes the selection procedure, if we assume that the wave functions are continuous. 

We illustrate the action of the genetic algorithm of state selection on our example with the association of protons in the molecular ion of hydrogen. We suppose that the initial state of the first proton and the atom of hydrogen are close sufficiently to the forming of molecular ion of hydrogen. We treat the electron not as the separate particle here, but as the factor which creates the attracting potential between protons. If the first choice of pairs (1 proton sample, 2 proton sample) is done such that many pairs were formed with the distance between protons far from the distance $r_0$ of stable molecular ion of hydrogen, then the protons from these pairs will fly to far distance one from another, and hence this pairs occur in the different groups. It results after several iterations in the growth of the groups where the initial distance between protons is close to $r_0$, and we obtain in the final the prevailing quantity of pairs which form the molecular ion of hydrogen. 

In case when the initial positions of the first proton and the hydrogen atom are far, this scheme has to be extended. Here we must take into consideration also photons, which are emitted by this system and thus decrease its energy. Photons can be included to the scheme with quantum state selection as well. But the direct consideration of photons is not necessary, because we can replace them by some kind of friction, and split the time frame $\Delta t$ to the smaller segments so that in the last segment the positions of flying proton and the target atom will be close enough to apply the selection procedure. 

The simulation we have proposed touches the basic things in quantum mechanics. At first it factually utilises the fundamental idea of many worlds (Everett). Each cortege consisting of samples of the real particles represent the separate quantum world. The different worlds interact with each other, and the selection process plays the role of judge in this interaction. The interaction inself results from the mechanism of dynamical diffusion when the close corteges exchange their impulses on the fixed place $j$. This exchange represents the exact form of the so called pre quantum fields considered in \cite{6}.

\section{Conclusion}

We propose the simple genetic algorithm for the simulation of association of atoms into molecules, based on the method of collective behavior. This algorithm is scalable in the sense that we can add new particles to the considered system and the construction still remains valid. In particular, it admits the inclusion of photons, and the complex molecules. The interesting peculiarity is that this algorithm can be reversed, the formal inversion, when we treat the dissociation process, gives us the picture of splitting of the molecule to the more simple molecules of atoms and we can estimate the probabilities of the both processes even for large quantity of participating atoms. 

The description of association and dissociation of molecule based on the method of collective behavior is completely quantum and is based on Hilbert formalism. This tool is scalable and can be applied to the complex chemical processes. In particular we can expect that this approach is applicable to the explanation of the phenomenon of molecular memory detected in experiments.
The proposed method for the simulation of quantum dynamics thus represents the kind of case technology in fundamental science.

\end{document}